# Self-powered Broadband Photodetector on Flexible Substrate from Visible to Near Infrared Wavelength

Hao Wang*[b], Chaobo Dong*[a], Yaliang Gui[a], Jiachi Ye[c], Salem Altaleb[c], Martin Thomaschewski[a], Behrouz Movahhed Nouri[a, b], Chandraman Patil[a], Hamed Dalir[c], Volker J Sorger[a,b]

[a]Department of Electrical and Computer Engineering the George Washington University, 800 22nd Street, Washington, DC 20052, USA

[b]Optelligence LLC, 10703 Marlboro Pike, Upper Marlboro, MD, 20772, USA

[c]Department of Electrical & Computer Engineering University of Florida, 968 Center Drive 216 Larsen Hall, Gainesville, FL 32611

* These authors contributed equally to this work.

## ABSTRACT

Van der Waals (vdWs) heterostructures assembled by stacking 2D crystal layers have proven to be a new material platform for high-performance optoelectronic applications such as thin film transistors, photodetectors, and emitters. Here, we demonstrate a novel device with strain tuning capabilities using $MoS_2/Sb_2Te_3$ vdWs p-n heterojunction devices designed for photodetection in the visible to near-infrared spectrum. The heterojunction devices exhibit remarkable characteristics, such as a low dark current in the range of a few picoamperes and a high photoresponsivity of 0.12 A $W^{-1}$. Furthermore, the proposed devices exhibit exceptional tunability when subjected to a compressive strain of up to 0.3%. By introducing strain at the interface of the heterojunction, the materials' bandgap is affected resulting in a significant change in the band structure of the heterojunction. This leads to a change in the detector's optical absorption characteristics improving the responsivity of the device. The proposed strain-induced engineering of the electronic and optical properties of the stacked 2D crystal materials allows tuning of the optoelectronic performance of vdWs devices for high-performance and low-power consumption applications for applications like wearable sensors and flexible electro-optic circuits.

## INTRODUCTION

Photodetectors are a crucial component in a wide range of applications, such as optical communication, sensing, imaging, and energy harvesting [1-3]. In recent years, there has been a growing interest in developing high-performance, flexible, and self-powered photodetectors for use in wearable intelligent sensors. Flexible photodetectors are designed to be bendable and conform to various shapes and curvatures, making them suitable for use in unconventional form factors such as smart textiles, medical devices, and electronic skin [4,5]. These devices have a wide range of potential applications, including health monitoring, military surveillance, and flexible displays, [6-9]. The development of flexible photodetectors requires the use of novel materials and device architectures that can withstand bending and stretching without compromising their electrical properties. Polymer films such as polyimide (PI), polyethylene naphthalate (PEN), or polyethylene terephthalate (PET) are common substrates used for flexible optoelectronics. Their mechanical properties, such as Young modulus and elastic strain are in the range of 2-7 GPa and 1-3.5 %, respectively [10-12]. Self-powered photodetectors, also known as photovoltaic photodetectors, can operate without external power sources. These devices work by converting light energy into electrical energy through the photovoltaic effect. Recent studies have shown that self-powered photodetectors based on 2D materials have several advantages over conventional device configurations, including higher sensitivity, lower power consumption, and smaller size, making them an attractive device platform for future optoelectronic technology [13, 14].

Transition metal dichalcogenides (TMDs) have been widely explored as active materials for high-performance optoelectronic devices due to their unique electronic and optical properties [15-17]. TMDs are layered materials consisting of two-dimensional (2D) layers of transition metal atoms (such as Mo, W, and Ti) sandwiched between chalcogen atoms (such as S, Se, and Te) in a hexagonal lattice structure [18]. The layered crystal structure results in unique electronic and optical properties that enable promising new applications. One of the critical properties of TMDs that makes them attractive for optoelectronics is their direct bandgap in the visible to near-infrared (NIR) range [19-20]. This direct bandgap allows TMDs to absorb light efficiently, making them suitable as active materials for photodetection. Another essential property of TMDs for optoelectronics is their high electron mobility, which enables fast charge transport and efficient device performance. The electron mobility in TMDs can range from a few hundred to several thousand $cm^2V^{-1}s^{-1}$. The hole mobility in TMDs can range from a few tens to several hundred $cm^2V^{-1}s^{-1}$ [21-23]. For example, $MoS_2$ (molybdenum disulfide), which is one of the most studied TMDs, has been shown to have electron mobility values in the range of $10-10^3$ $cm^2V^{-1}s^{-1}$ and hole mobility values in the range of 20-200 $cm^2V^{-1}s^{-1}$ [21, 24]. TMDs also exhibit significant nonlinear coefficients due to their 2D layered structure, which enhances their nonlinear light-matter interactions, including photoluminescence and Raman scattering [25-27]. In addition, the atomically thin nature of TMDs allows for the fabrication of flexible and transparent devices, making them suitable for wearable and flexible optoelectronics [10, 28, 29]. However, TMDs are frequently impeded by surface oxidation, high contact resistance, and low mobilities [30-32]. To overcome these shortcomings in TMDs-based photodetectors, recent advancements have been made utilizing techniques such as plasmonics to enhance the light-matter interaction of TMDs, and heterostructures to modify their intrinsic electrical and optical properties as the built-in electric field can efficiently separate the electron-hole pairs. These methods have demonstrated significant performance improvements, including improved sensitivity, broader frequency response, enhanced external quantum efficiencies (EQE), heightened detectivity (D*), and a more extensive spectral photoresponse [20, 23, 33-35]. The combination of transition metal dichalcogenides with other emerging materials, such as topological insulators (TIs), has not been widely explored yet. TIs possess several attractive optoelectronic properties, such as a unique energy band structure or topologically protected conductive edge or surface states, which offer higher mobility and a broader detection spectrum than graphene, which lacks a bandgap. The low thermal conductivity makes it beneficial to reduce heat dissipation and improve the efficiency of photovoltaic devices. In addition, TIs are topologically protected, which means that their surface states are robust against perturbations and immune to scattering by non-magnetic impurities or defects. [36-40] This makes TIs highly stable and durable, which is vital for the development of practical optoelectronics devices.

Molybdenum disulfide ($MoS_2$), as one of the most representative TMDs, has a direct bandgap of 1.8 eV, whereas, in bulk, it has an indirect bandgap of 1.3 eV. It can be used in various optoelectronic applications such as photovoltaics and photodetection due to its unique optical and electronic properties, including a direct bandgap, high absorption coefficient, and absorption spectrum extending from visible to infrared [41–44]. Antimony telluride ($Sb_2Te_3$) is a topological insulator. Its bandgap is approximately 0.2 to 0.3 eV, which places it in the category of a narrow-bandgap semiconductor. It exhibits a topologically protected surface state that can lead to high carrier mobility and low noise [45-47]. Heterojunctions are widely applied for self-powered photodetectors due to the built-in electric field that can efficiently separate the electron-hole pairs to enhance the photoresponsivity and speed under zero bias. By combining $Sb_2Te_3$ and $MoS_2$ in a heterojunction structure, it is possible to achieve complementary optoelectronic properties and enhance the performance of the resulting photodetector. Previous studies have demonstrated the potential of $Sb_2Te_3/MoS_2$ heterojunction-based photodetectors. Wang et al. demonstrated a self-powered photodetector based on $Sb_2Te_3/MoS_2$ heterojunction that covered a wavelength range from the visible to the near-infrared (NIR). The device exhibits a low dark current of 2.4 pA at zero bias and high photoresponsivity of >150 mA $W^{−1}$ at zero bias [42]. Liu et al. reported a photodetector based on a $Sb_2Te_3/MoS_2$ heterojunction that exhibits a high photoresponsivity of 130 A/W and a detectivity of $2.2 \times 10^{12}$ Jones [48]. However, these previous studies did not address the challenges of developing flexible and self-powered $Sb_2Te_3/MoS_2$ heterojunction-based photodetectors, as the structural flexibility of the device is critical for applications such as wearable electronics and conformal sensors. Here, we demonstrate a self-powered broadband photodetector based on a $Sb_2Te_3/MoS_2$ heterojunction on the PI substrate. The photodetector exhibits excellent performance with a detection wavelength ranging from visible to NIR. For example, it shows an ultra-low dark current of 4.3 pA and a high photoresponsivity of 120 mA W-1. Additionally, the proposed device demonstrates outstanding stability when subjected to compressive strain up to 0.3%. This phenomenon opens up new possibilities for innovative, flexible optoelectronic devices which can harness the unique properties of heterojunctions for enhanced performance and functionality. [49-61]

**RESULTS**

The heterojunction of MoS$_2$ and Sb2Te$_3$ flakes is fabricated by mechanically exfoliating the flakes from their respective bulk crystals. The flakes are carefully transferred onto a clean, flexible polyimide thin film in a specific sequence, using a 2D transfer system that minimizes contamination of material surfaces and avoids potential structural damage to the flakes. The device schematic is illustrated, providing a visual representation of the different components of the heterojunction and their arrangement (Fig.1a). Microscope image of the fabricated MoS$_2$/Sb$_2$Te$_3$ can be seen as depicted in Fig.1b where the Sb$_2$Te$_3$ flake is transferred on top. The electrodes were formed of Ti/Au (5/45 nm) both the MoS$_2$ and Sb$_2$Te$_3$ flakes for electrical contacts. The deposition of the gold electrodes ensures a good electrical connection and minimizes parasitic effects that could affect the performance of the heterojunction for the chosen materials here. To study the impact of strain on the p–n heterojunction, the flexible device was bent upwards to apply tensile strains. The applied strain was calculated based on the introduced bending angles, which were carefully measured using a customized tensile stress setup. The application of tensile strains allows for the investigation of the impact of mechanical deformation on the electronic and optical properties of the heterojunction. Finally, the flexible heterojunction was characterized under different wavelengths and intensities of illumination using a supercontinuum laser source. The detailed characterization process includes measuring the electrical and optical properties of the heterojunction as a function of the applied strain. The results of this study could have significant implications for the development of flexible and high-performance electronic devices.

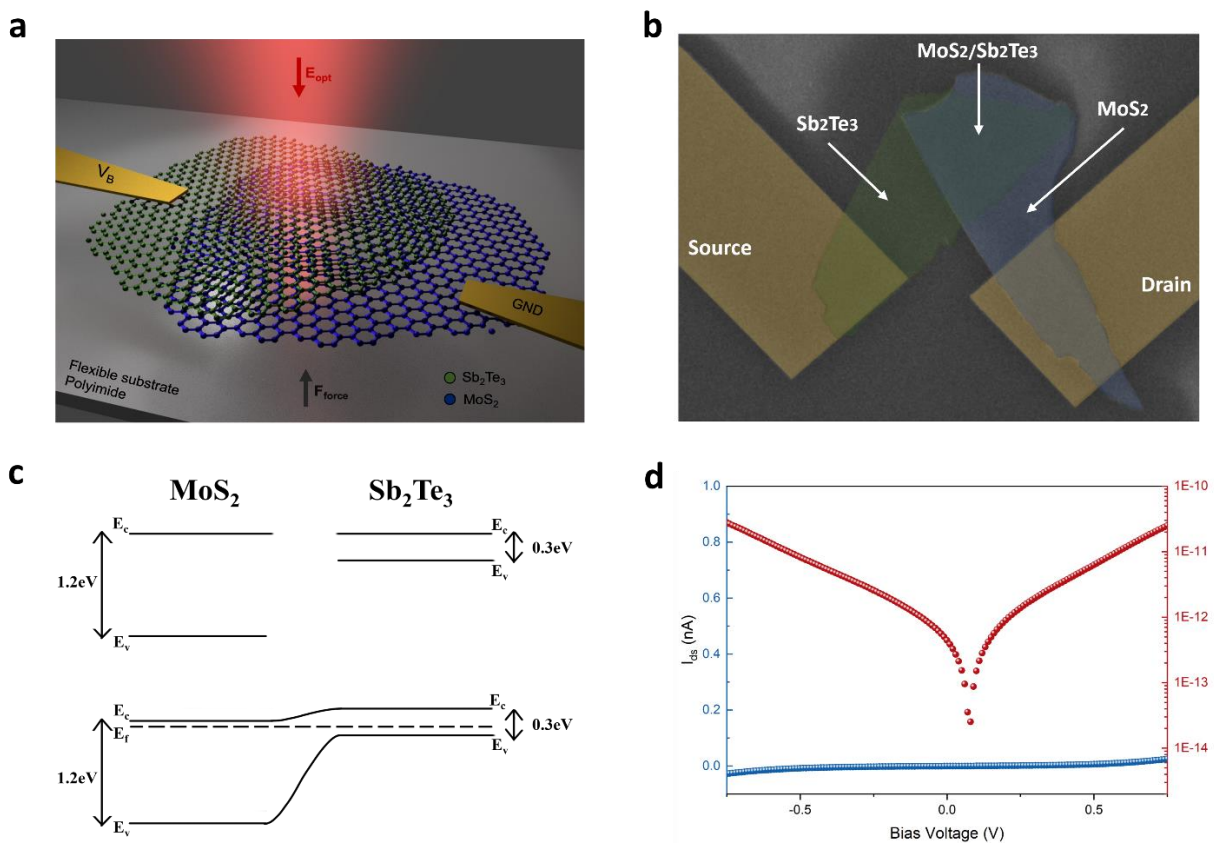

Figure 1. Sb$_2$Te$_3$/MoS$_2$ heterostructure photodetector (a) A schematic representation of the Sb$_2$Te$_3$/MoS$_2$ van der Waals heterojunction photodetector. (b) The SEM image of the Sb$_2$Te$_3$/MoS$_2$ heterostructure device. (c) The band structures of the vdW layered MoS$_2$/Sb$_2$Te$_3$ heterojunction. (d) The I–V characteristics of the Sb$_2$Te$_3$/MoS$_2$ van der Waals heterojunction under -0.75 to 0.75V bias voltage in a dark environment.

Electron-hole pairs are generated at the junction upon incident optical illumination on the material, which can be measured as an electrical current under electrical bias. Nonetheless, a p-n heterojunction can even detect light in the absence of voltage due to the presence of built-in junction potential. This phenomenon results from the alignment of energy levels in both materials (Fig. 1c). The built-in potential generates an electric field across the junction for electrons and holes to be collected at the electrodes. Before conducting the optoelectronic measurements, we performed an assessment of the electrical properties of the $MoS_2/Sb_2Te_3$ junction by applying a source-drain bias voltage ($V_{sd}$) ranging from -0.75 V to 0.75 V. This evaluation allowed us to investigate the dark current behavior and assess the performance of the van der Waals p-n heterojunction. As shown in Fig.1d, the dark current remained stable and consistent across the range of applied bias voltages. Specifically, the measured dark current was found to be as low as a few picoamperes (minimum system noise floor) at zero bias and increased merely to 29 pA / 23 pA at ±1 V bias. These low levels of dark current are a direct result of the unique heterostructure design, which effectively suppresses unwanted leakage currents and improves device performance enabling high noise equivalent power (NEP).

To evaluate the spectral response of the photodetectors, the devices are illuminated with a supercontinuum laser source. A wavelength sweep on the illuminated device was performed to observe the photocurrent for the spectrum. The light is focused onto the junction area of the photodetectors using the different power levels 0.55 uW, 2.01 uW, and 3.85uW, respectively.

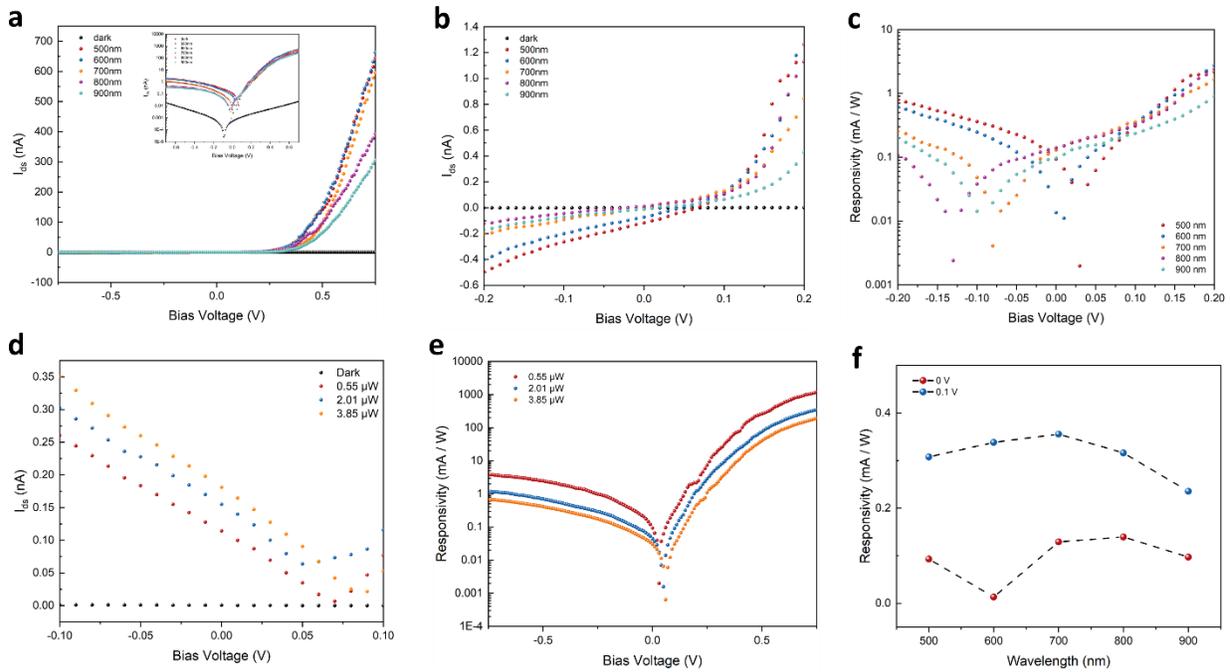

Figure 2. Photovoltaic characteristics of the $Sb_2Te_3/MoS_2$ van der Waals heterojunction. (a), (b) The I–V characteristics of the $Sb_2Te_3/MoS_2$ heterojunction under different wavelength optical with the same illumination power of 0.55 uW. (c) The measured photoresponsivity at different wavelengths, 500 nm, 600nm, 700 nm, 800nm, and 900nm, with the same illumination power of 0.55 uW, under -0.2 to 0.2 V bias voltage. (d) The measured photocurrent at 500 nm in three different power levels: 0.55 uW, 2.01 uW, and 3.85uW, respectively. (e) The measured photoresponsivity at 500 nm in three different power levels: 0.55 uW, 2.01 uW, and 3.85uW, respectively. Photoresponsivity was observed and increased proportionally to the optical power, under -0.75 to 0.75 V bias voltage. (f) Normalized responsivity by sweeping the wavelength under 0 and 0.1V bias from 500 to 900 nm with a step size of 100 nm, which indicates a broad wavelength response.

The power-dependent current-voltage (IV) characteristics measurement of the p-n heterojunction has been conducted across a range of wavelengths from 500 nm to 900 nm with a step size of 100 nm. When the bias voltage is zero, the heterojunction effectively separates the photogenerated electron-hole pairs, which then generate a photocurrent due to the built-in electric field established at the $MoS_2/Sb_2Te_3$ interface. In Fig. 2a and 2b, the relationship between the measured photocurrent ($I_{photo}$) and the wavelength can be observed. The optical power remains the same at 0.55 uW for all the corresponding wavelengths. When the wavelength of the incident light is shifted to higher energies (shorter wavelengths), the photons carry more energy and can excite electrons to higher energy levels in the conduction band. This leads to an increase in the photocurrent generated. The shift of the photocurrent dip indicates the change in the energy of the incident light. The depth of the dip depends on the specific properties of the semiconductor material, such as the bandgap energy and the density of states in the conduction and valence bands. As the intensity of the optical power illumination increases, a corresponding rise in the photocurrent is observed due to the amplified generation rates of the electron-hole pairs (Fig.2d). This behavior emphasizes the potential for self-powered photodetection across a broad range of wavelengths. The relationship between the incident optical power and the resulting photocurrent is often nonlinear and can be affected by factors such as the bandgap energy of the semiconductor material, the doping level, and the contact configuration of the device. However, although the increase in incident optical power will lead to a corresponding increase in photocurrent, after reaching a certain power intensity, the saturation effect or the damage of the device will undermine the responsivity performance (Fig.2e). This saturation effect highlights the importance of understanding the optimal operating conditions for the photodetection device, as it may influence the device's overall efficiency and effectiveness in various applications. Balancing the illumination intensity and other parameters is crucial to maximizing the device's responsivity without reaching the saturation threshold, ensuring optimal performance in a wide range of photodetection scenarios.

As depicted in Fig.2f, the device exhibits a photoresponsivity of approximately 0.1 mA/W in the absence of any external bias. This finding underscores the heterojunction's ability to effectively perform photodetection tasks without the need for additional bias voltage, highlighting its potential for ultra-low power applications. Measuring the wavelength-dependent responsivity of the photodetection device, we found that, under a bias voltage of 0.1 V, the difference in responsivity from 500 nm to 900 nm wavelengths is less than 0.1 mA/W. This observation suggests that the proposed device is capable of functioning effectively across a broad photoresponsive spectrum, ranging from visible to near-infrared (NIR) regions. Moreover, the device exhibits a non-dispersive responsivity of approximately 0.3 mA/W when operating under a 0.1 V bias, demonstrating its potential for efficient broadband photodetection applications with constant responsivity over a broad wavelength range. When a positive bias voltage is applied, an external electric field is created at the junction interface, which enhances the carrier separation efficiency of the photogenerated carriers, consequently boosting the responsivity. At a +1 V bias, the responsivity experiences an amplification depending on both the wavelength and the optical power. As the applied voltage increases, the responsivity continues to grow and eventually reaches a saturation point that is contingent on the incident optical power.

To gain a deeper insight into the impact of mechanical strain on photoresponsivity, an experiment to measure the photocurrent ($I_{ds}$) at varying levels of strain and light intensity was conducted. Initially, the I-V curves were measured under different levels of strain (0 %, 0.12 %, and 0.3 %) in the absence of light, as shown in Fig.3a.b. We observe a significant strain-modulated behavior, with the rectification characteristics weakening under tensile strain conditions. These findings suggest that the application of mechanical strain can have a significant impact on the electrical properties of the heterojunction, leading to changes in the dark current. At a bias voltage of 0.6 V, we observed a decrease in the dark current from 14 pA to 9 pA as the tensile strain increased from 0% to 0.12%, with a further drop to 2 pA at a strain of 0.3%. This behavior can be attributed to the strain-induced changes in the device's energy band structure. However, the signal became noisier as the current levels approached the measurement setup's limits, with environmental noise being a significant factor. It is important to carefully control the testing environment and to use appropriate measures to reduce any sources of noise that could affect the results.

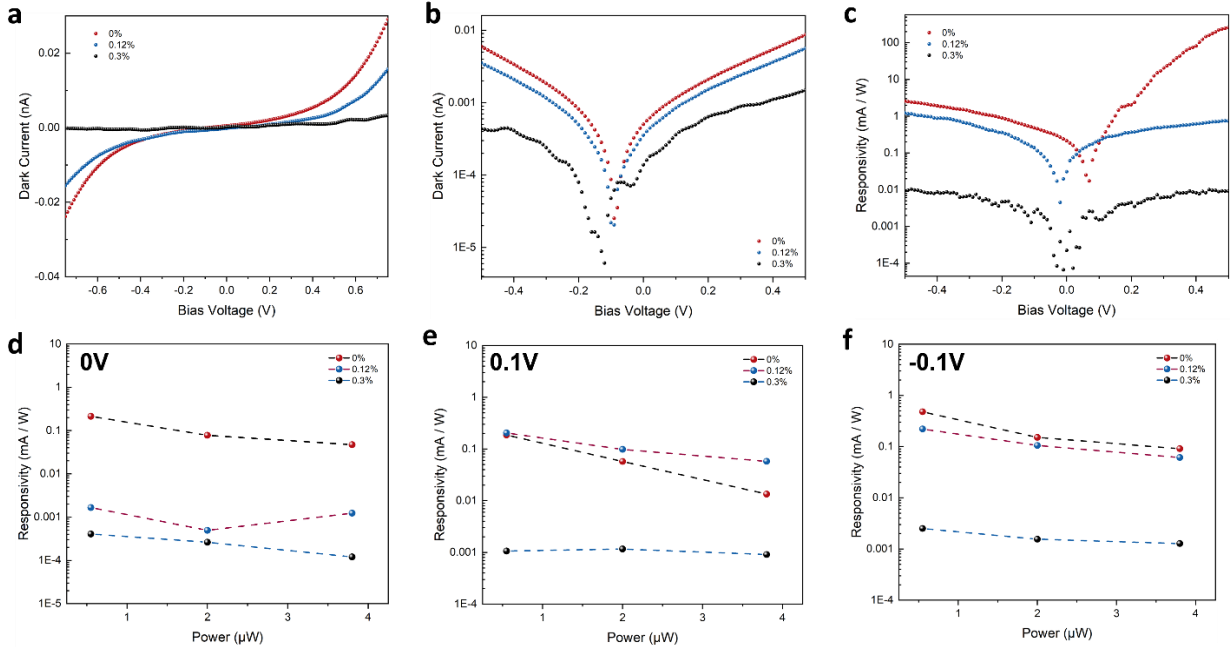

Figure 3. Strain-dependent photoresponse of the $Sb_2Te_3$/$MoS_2$ van der Waals heterojunction. (a), (b) The I-V characteristic of the device was measured under varying strains without illumination on a flexible substrate. The dark current decreased substantially as the tensile strain increased. (c) The photoresponsivity of the device was measured under different tensile strains at 500nm wavelength illumination. As the strain increased, the responsivity decreased, indicating a simultaneous drop in both the photo and dark currents. However, the photocurrent exhibited a more significant decrease. (d), (e), (f) The power-dependent responsivity change of the heterojunction device was measured under different strains, with a 500nm laser illumination at 0V, 0.1V, and -0.1V. The results indicated that as the voltage increased, the photoresponsivity decreased, indicating that the photocurrent had reached

Mechanical strain can also introduce changes in the resistance of a photodetection device. When a semiconductor material is subjected to mechanical strain, the lattice structure of the material is distorted, leading to changes in the effective cross-sectional area and length of the material. As a result, the resistance of the device is altered due to changes in the free carrier concentration and mobility. Under tensile strain conditions, the resistance of the device generally increases due to the reduction in the doping concentration and carrier mobility. This increase in resistance can have a significant impact on the performance of the photodetection device, as it can affect the magnitude of the photocurrent and the photoresponsivity of the device. Furthermore, the change in resistance can also impact the noise level of the device. The increase in resistance can lead to an increase in the Johnson noise of the device, which is generated due to the thermal agitation of the carriers in the material. This noise can contribute to the overall noise level of the device and reduce its signal-to-noise ratio. Compared to the zero-strain condition, the separation of electron-hole pairs by the built-in electric field occurs primarily in or near the heterojunction upon illumination, with electrons (holes) being swept into the $MoS_2$/$Sb_2Te_3$ layer, resulting in a photocurrent generation. When a tensile strain is applied, the total internal electric field and the electric potential difference within the heterojunction are reduced, as previously discussed. This reduction hinders the separation of electron-hole pairs and impairs the injection efficiency in or near the heterojunction, leading to a decrease in both $I_{ph}$ and responsivity when the tensile strain increases.

## Conclusion

In conclusion, mechanical strain significantly affects the performance of a photodetection device, with strain-induced changes in the bandgap, resistance, and depletion region all affecting the device's photoresponsivity. Under conditions of tensile strain, the bandgap and resistance increase while the doping concentration and carrier mobility decrease, resulting in a decrease in photoresponsivity. The results have significant implications for the advancement of photodetection devices, particularly in applications where mechanical strain may be present, such as wearable devices and flexible electronics. By comprehending the effects of mechanical strain on the performance of photodetection devices, it may be possible to design more robust and reliable devices that can function effectively under a variety of strain conditions.

## Methods

The tunable (NKT SUPERCONTINUUM Compact) source and the source meter (Keithly 2600B) were used for electrical response measurements of $Sb_2Te_3$/$MoS_2$ PN junction heterostructure devices. The laser beam was focused on the devices by an objective lens.

**Acknowledgments**

**Author contributions**

H.W. designed the study. C.D. fabricated the device, also built the experimental system and performed the experiments with H.D., H.W. and V.J. supervised the project. All of the authors wrote and revised the manuscript.